\documentclass[sigconf]{acmart}

\usepackage{amsmath}

\usepackage{microtype,xspace,graphicx,fancyvrb,multirow}
\usepackage{listings}
\usepackage{soul} 
\usepackage{enumitem}

\input{pygcode/friendly.pygstyle}

\makeatletter
\providecommand{\@LN@col}[1]{}
\makeatother

\usepackage{float}
\usepackage{algorithm2e}
\usepackage{fontawesome5}

\makeatletter
\let\old@lstKV@SwitchCases\lstKV@SwitchCases
\def\lstKV@SwitchCases#1#2#3{}
\makeatother
\makeatletter
\let\lstKV@SwitchCases\old@lstKV@SwitchCases

\lst@Key{numbers}{none}{%
    \def\lst@PlaceNumber{\lst@linebgrd}%
    \lstKV@SwitchCases{#1}%
    {none:\\%
     left:\def\lst@PlaceNumber{\llap{\normalfont
                \lst@numberstyle{\thelstnumber}\kern\lst@numbersep}\lst@linebgrd}\\%
     right:\def\lst@PlaceNumber{\rlap{\normalfont
                \kern\linewidth \kern\lst@numbersep
                \lst@numberstyle{\thelstnumber}}\lst@linebgrd}%
    }{\PackageError{Listings}{Numbers #1 unknown}\@ehc}}
\makeatother

\newcommand{\cc}[1]{{\smaller[0.5]\texttt{#1}}}

\newcommand{\figrule}{\hrule width \hsize height .33pt}
\newcommand{\coderule}{\vspace{0.2em}\figrule\vspace{0.2em}} 

\fvset{fontsize=\footnotesize,xleftmargin=8pt,numbers=left,numbersep=5pt}
\input{code/fmt}

\definecolor{editorGreen}{rgb}{0, 0.5, 0}
\definecolor{lightgray}{rgb}{0.95, 0.95, 0.95}
\definecolor{LightCyan}{rgb}{0.88,1,1}
\definecolor{lightblue}{HTML}{C7DCF0}
\newcommand{\codeknow}[1]{\sethlcolor{lightblue}\hl{#1}}
\definecolor{lightpurple}{HTML}{CCCBFF}
\newcommand{\lemmaknow}[1]{\sethlcolor{lightpurple}\hl{#1}}
\definecolor{lightyellow}{HTML}{FDF2CF}
\newcommand{\verusknow}[1]{\sethlcolor{lightyellow}\hl{#1}}

\newcommand{\setlistcolorline}[2]{\ifnum\value{lstnumber}=#1\color{#2}\fi}
\newcommand{\setlistcolorlines}[3]{\ifnum\value{lstnumber}>#1\ifnum\value{lstnumber}<#2\color{#3}\fi\fi}

\definecolor{editorOcher}{rgb}{1, 0.5, 0} 
\definecolor{lightgray}{rgb}{0.95, 0.95, 0.95}

\newcommand{\ourtool}{\mbox{\textsc{KVerus}}\xspace}

\newcommand{\autoverus}{\mbox{\textsc{AutoVerus}}\xspace}
\newcommand{\safe}{\mbox{SAFE}\xspace}

\newcommand{\alphaverus}{AlphaVerus\xspace}

\newcommand{\laurel}{\mbox{\textsc{LAUREL}}\xspace}

\newcommand{\preproc}{\textit{Preprocessor}\xspace}
\newcommand{\comp}{\textit{Comprehender}\xspace}
\newcommand{\prv}{\textit{Prover}\xspace}
\newcommand{\san}{\textit{Refiner}\xspace}

\begin{document}
\date{}

\title{\ourtool: Scalable and Resilient Formal Verification Proof Generation for Rust Code}

\author{Yuwei Liu}
\correspondingauthor
\orcid{0000-0001-5170-3388}
\affiliation{%
  \institution{Ant Group}
  \city{Hangzhou}
  \country{China}
}
\email{lyw458372@antgroup.com}

\author{Xinyi Wan}
\orcid{0009-0004-3543-499X}
\affiliation{%
  \institution{Ant Group}
  \city{Shanghai}
  \country{China}
}
\email{wanxinyi.wxy@antgroup.com}

\author{Yanhao Wang}
\orcid{0000-0002-6990-2972}
\affiliation{%
  \institution{Independent}
\city{Beijing}
  \country{China}
}
\email{wangyanhao136@gmail.com}

\author{Minghua Wang}
\orcid{0000-0002-2270-2076}
\affiliation{%
  \institution{Ant Group}
  \city{Beijing}
  \country{China}
}
\email{minghua.wmh@antgroup.com}

\author{Lin Huang}
\orcid{0009-0002-5659-1471}
\affiliation{%
  \institution{Ant Group}
  \city{Beijing}
  \country{China}
}
\email{linyu.hl@antgroup.com}

\author{Tao Wei}
\orcid{0000-0001-9537-7051}
\affiliation{%
  \institution{Ant Group}
  \city{Hangzhou}
  \country{China}
}
\email{lenx.wei@antgroup.com}

\ccsdesc[500]{Software and its engineering~Formal software verification}

\keywords{Formal Verification, Large Language Models, Proof Generation, Rust, Verus}

\begin{abstract}
Repository-scale formal verification is becoming practical for Rust systems, but maintaining and extending proofs remains expensive in real projects. 
In the Asterinas Rust OS kernel, the CortenMM memory-management module requires Verus proofs that cross file and module boundaries, reuse project-specific lemmas with limited documentation, and survive frequent Verus/toolchain evolution.
Existing LLM-based proof-generation methods are mostly designed for small or single-file tasks, and they do not provide the structural project knowledge or verifier-aware repair needed in such a repository-scale verification workflow.

This paper presents \ourtool, a practical LLM-assisted workflow for Verus proof generation in evolving Rust repositories.
\ourtool builds a dynamic knowledge base over code metadata, lemma semantics, and Verus/toolchain knowledge.
It combines dependency-aware context extraction, semantic lemma retrieval, and error-driven refinement to generate Verus proof code that is checked by the verifier and then reviewed by engineers.
In CortenMM, \ourtool generated verifier-passing proof patches that verified 23 previously unverified functions, introduced 6 reusable lemmas, and accounted for 21.0\% of the proof code in the module; these patches were accepted upstream by Asterinas developers.
The CortenMM experience also exposed practical boundary cases, including incorrect specifications that required human repair and generated proofs that benefited from human cleanup.
Across benchmarks, \ourtool achieves a 51.0\% success rate on three repository-level benchmarks with cross-file dependencies, compared with 4.5\% for a multi-round prompting baseline, and verifies 80.2\% of tasks on three single-file benchmarks.
These experiences and results show that LLM-assisted proof generation can move beyond benchmark-oriented synthesis and become a maintainable workflow for repository-scale Rust verification.
\end{abstract}

\maketitle

\section{Introduction}
\label{sec:intro}

Rust is increasingly used in systems software where memory safety, concurrency control, and low-level resource management are critical.
Formal verification has already delivered strong assurance for kernels, compilers, and system infrastructure~\cite{klein2009sel4,gu2016certikos,leroy2016compcert}, but applying it to a real, evolving repository remains labor-intensive.
This challenge is concrete in Asterinas~\cite{peng2025asterinas}, a general-purpose Rust OS kernel, and its verified memory-management module CortenMM~\cite{Zhang2025CortenMM}.
Proof engineers working on CortenMM must construct Verus~\cite{lattuada2023verus,lattuada2024verus} proofs that depend on specifications, traits, type aliases, helper functions, and lemmas distributed across the repository.
Unlike interactive theorem proving environments such as Coq/Rocq~\cite{bertot2013interactive} or annotation-oriented verifiers for other languages such as VST-A~\cite{zhou2024vst}, repository-scale Verus verification must align Rust code, ghost proof code, SMT automation, and fast-changing toolchain behavior.
They must also maintain these proofs as the implementation and the Verus toolchain evolve.
Over six months, the CortenMM verification code required eight synchronizations with Verus, averaging roughly one synchronization every three weeks.
In day-to-day verification work, these costs appear in several recurring forms.
Before writing a proof, an engineer must identify which local specifications, page-table helpers, arithmetic lemmas, and state-machine invariants are relevant to the target function.
Many of these dependencies are not documented as natural-language proof hints; their meaning must be inferred from Verus signatures, preconditions, postconditions, and existing proof bodies.
After a candidate proof is written, the engineer must interpret verifier diagnostics, repair missing assertions or lemma invocations, and ensure that the final proof is concise and maintainable enough to be reviewed by the kernel developers.
In this setting, proof generation is not a stand-alone coding task; it is part of an engineering workflow that must discover project context, reuse existing proof knowledge, pass the verifier, and produce code suitable for human review and upstream acceptance.

Large Language Models (LLMs) offer a promising way to reduce this proof burden.
Recent work has shown that LLMs can synthesize proof annotations or proof code for small verification tasks~\cite{ma2024specgen,thompson2024rango,chen2024automated,aggarwal2024alphaverus,yang2024autoverus,lu2024proof,Zhang2026VerusSeek}.
However, these methods are largely benchmark-oriented and often assume a small, self-contained program or a fixed prompt with manually curated examples.
That assumption breaks down in repository-scale Verus verification.
A target proof may require definitions several files away; the most useful lemma may be project-specific and undocumented; and a Verus syntax or library change may invalidate previously successful proof patterns.
Simply enlarging the prompt is not a reliable solution, because irrelevant repository code consumes context, distracts lemma selection, and still leaves the model without an explicit account of which dependencies are structurally required by the verifier.
Likewise, repeated prompting without verifier-specific knowledge often repairs local syntax errors while missing the underlying proof obligation, such as a required arithmetic fact, a trait-bound implication, or a state-machine invariant.
For example, Verus-Bench, a suite derived from \autoverus, was significantly affected by recent syntax changes to loop constructs, and more than 400 lines had to be revised to make the suite compatible with the latest Verus release.
These failures are not just prompt-quality issues: they reflect a mismatch between semantic code-pattern generation and the structural constraints imposed by a verifier and a modular codebase.

We call this mismatch the \emph{Semantic-Structural Gap}.
LLMs operate over semantic regularities in code and text, whereas formal verification is governed by precise dependency graphs, logical preconditions and postconditions, lemma availability, verifier syntax, and toolchain-specific diagnostics.
In a repository-scale verification workflow, bridging this gap requires more than asking an LLM to produce a proof once.
The proof-generation system must expose the repository structure relevant to a target proof, retrieve reusable proof knowledge even when it is not well documented, and refine failed attempts using verifier feedback and version-aligned toolchain knowledge.

This paper presents \ourtool, a practical LLM-assisted workflow for Verus proof generation in evolving Rust repositories.
Given an unverified Verus function, \ourtool first extracts dependency-scoped code context from repository metadata, including relevant functions, specifications, traits, structures, and type aliases.
It then retrieves reusable lemmas from both project code and Verus libraries by indexing lemma signatures and LLM-generated semantic summaries.
Finally, \ourtool generates candidate proof code, checks it with Verus, and uses verifier diagnostics together with Verus/toolchain knowledge to refine failed attempts.
The resulting workflow is designed to fit the path used in real verification work: unverified repository code, automated proof generation, Verus checking, error-driven refinement, human review, and upstream-accepted proof code.
\ourtool therefore does not treat the LLM output as a trusted proof artifact.
The verifier remains the authority for proof validity, and human review remains the authority for whether the generated proof should be integrated into the kernel repository.
This separation is important for practical use: automation reduces repetitive proof construction and maintenance work, while the existing Verus and code-review process continues to guard correctness and maintainability.

We report both controlled experiments and practical verification experience with \ourtool.
First, we applied \ourtool to CortenMM in the Asterinas Rust OS kernel.
\ourtool generated 1,387 lines of proof code, including 6 newly introduced lemma functions, and verified 23 functions that were previously unverified.
The generated code accounts for 21.0\% of the proof code in CortenMM and was reviewed and accepted by Asterinas developers.
These proofs cover realistic verification tasks, including nonlinear arithmetic, state-machine invariants, and bit-level executable code.
This result shows that \ourtool can be integrated into a real Rust OS verification workflow rather than only solving isolated benchmark tasks.
It also shows that repository-specific knowledge is not merely an optimization for benchmark accuracy: it is necessary for producing proof code that survives the complete path from generation to verifier acceptance and upstream review.

Second, we evaluate whether the workflow generalizes beyond this Asterinas/CortenMM experience.
On three repository-level benchmarks, \ourtool achieves a 51.0\% success rate on tasks with cross-file dependencies.
A multi-round prompting baseline achieves 4.5\%.
On three single-file benchmarks, \ourtool verifies 80.2\% of tasks, outperforming the state-of-the-art \autoverus (56.9\%) while reducing token cost by more than 50\%.
\ourtool also degrades less than \autoverus under breaking Verus updates, indicating that toolchain evolution must be treated as a first-class maintenance concern in long-lived Verus repositories.

This paper makes three contributions:

\begin{itemize}
    \item \textbf{Practical verification workflow.} We present \ourtool, a verifier-in-the-loop and human-review-centered workflow for generating Verus proof code in evolving Rust repositories. \ourtool combines dependency-aware context extraction, semantic lemma retrieval, and error-driven refinement while preserving Verus checking and human review as mandatory validation steps.
    \item \textbf{Open-source verification experience.} We report our experience using \ourtool on CortenMM, Asterinas's memory-management module. \ourtool generated verifier-passing proof patches that verified 23 previously unverified functions, introduced 6 reusable lemmas, and were accepted upstream by Asterinas developers.
    \item \textbf{Engineering assessment and lessons.} We analyze the CortenMM verification experience, including accepted proof patches, generated proof verbosity, and boundary cases involving incorrect specifications and human cleanup. We distill lessons for building LLM-assisted proof-generation workflows that remain maintainable under repository and verifier evolution.
\end{itemize}

\section{Open-Source Verification Context and Motivation}
\label{sec:bg}

\subsection{Verus-Based Rust Verification Workflow}

Verus~\cite{lattuada2023verus,lattuada2024verus} is a Rust-native deductive verifier that lets developers write specifications and proof code alongside Rust programs.
It is part of a broader Rust verification landscape that includes deductive verifiers, model checkers, interpreters, and static analyzers~\cite{astrauskas2022prusti,denis2022creusot,vanhattum2022kani,miri,mirai,wang2024unsafecop}, as well as ongoing efforts to verify core Rust libraries~\cite{verifyruststd}.
For \ourtool, the relevant Verus concepts are compact: a target function or lemma is annotated with preconditions, postconditions, invariants, and ghost proof steps; Verus translates these annotations into verification conditions and discharges them using an SMT solver such as Z3~\cite{de2008z3}.
When automation fails, developers usually add assertions, invoke helper lemmas, strengthen invariants, or adjust proof structure until the verifier accepts the code.

CortenMM~\cite{Zhang2025CortenMM}, the verified memory-management module of the Asterinas Rust OS kernel~\cite{peng2025asterinas}, illustrates why this workflow becomes difficult at repository scale.
A proof target may depend on specifications, traits, type aliases, page-table helper functions, and previously proven lemmas spread across multiple files.
The proof engineer must first identify the relevant repository context, then select reusable lemmas, then repair verifier failures caused by missing assertions, solver-boundary issues, trigger instability~\cite{leino2016trigger}, or Verus/toolchain changes.
Thus, the practical problem is not merely generating a few lines of proof code for a self-contained example; it is supporting the complete path from an unverified repository function to verifier-passing proof code that can be reviewed and accepted upstream.

\subsection{Motivating Example}

\autoref{code:example-proof} shows the \ourtool-generated proof for an unproven Verus lemma from CortenMM.
The target lemma, \cc{lemma\_va\_range\_get} \cc{\_tree\_path}, proves properties of the page-table traversal path derived from a virtual-address range \cc{va}.
Its precondition requires the range to satisfy \cc{va\_range\_wf}, and its postconditions require every node in the returned path to be a valid page-table node and require the path length to match the guard level of the address range.
Even understanding the proof obligation requires context beyond the local file: the target depends on address-range specifications, cursor/page-table helper functions, \cc{NodeHelper} specifications, type aliases, and arithmetic lemmas.

Existing LLM-based Verus proof-generation tools are not designed for this setting.
Single-file prompting misses the definitions needed to understand the target, while repository-wide prompting includes too much irrelevant code and still does not identify which dependencies are structurally required.
Moreover, useful project-specific lemmas are often undocumented, so a proof generator must reason from formal signatures and postconditions rather than from natural-language explanations.
The important point is not only that proof code is produced, but that the generated proof uses dependency-scoped context and reusable lemmas from outside the target file.
We follow this target through dependency extraction, lemma retrieval, and prompt construction in \autoref{sec:preproc} and \autoref{sec:prover}.

\begin{figure}[t]
    \centering
    \input{pygcode/example-proof.pygtex}
    \coderule
    \begin{lstlisting}[
    caption={\textbf{Generated proof for a CortenMM lemma.} The highlighted code is generated by \ourtool, while comments indicate the source of external dependencies required for verification.}, label={code:example-proof}, linewidth=\linewidth]
    \end{lstlisting}
\end{figure}

\subsection{Practical Challenges in Repository-Scale Proof Generation}
\label{sec:bg:challenges}

\textbf{Challenge 1: Cross-module dependency discovery.}
A target Verus proof may rely on specifications, helper functions, traits, structures, type aliases, and lemmas located several files away.
The generator must retrieve enough context for the verifier-relevant proof obligation without flooding the prompt with unrelated repository code.

\noindent\textbf{Challenge 2: Project-specific lemma reuse.}
Repository-scale proofs depend heavily on reusable lemmas, but project lemmas often lack natural-language comments or complete documentation.
The generator must infer lemma meaning from signatures, \cc{requires}/\cc{ensures} clauses, module locations, and existing proof structure, then retrieve lemmas that match the current proof obligation.

\noindent\textbf{Challenge 3: Verus/toolchain evolution.}
Verus is actively evolving, and changes to syntax, libraries, or verifier behavior can break previously valid proof patterns.
For long-lived repositories such as CortenMM, proof generation must therefore include version-aligned Verus knowledge and verifier-aware repair, rather than relying on static examples embedded in a prompt.

\subsection{Our Solution}
\label{sec:solution}

\ourtool addresses these challenges with a knowledge-centric proof-generation workflow.
It first extracts dependency-scoped code context from repository metadata, then retrieves reusable project and Verus-library lemmas through semantic indexing, and finally generates and refines proof code using Verus diagnostics.
The workflow keeps Verus and human review in the loop: generated proof code must pass the verifier, and upstream integration remains subject to normal repository review.
The next section describes this workflow in detail.

\section{\ourtool Architecture and Workflow}
\label{sec:design}

\begin{figure*}[t]
    \centering
    \includegraphics[width=.95\textwidth]{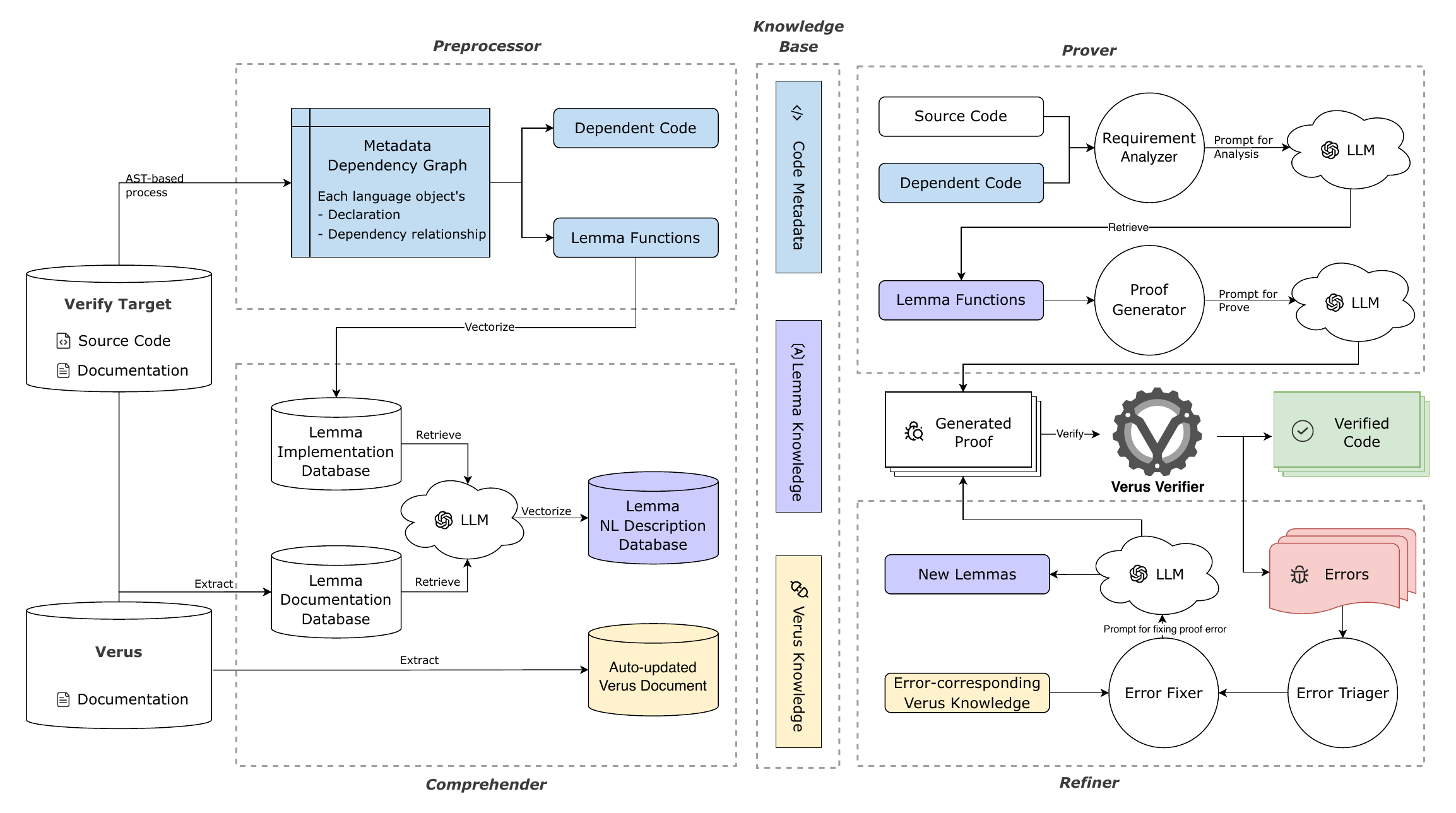}
    \caption{\textbf{\ourtool architecture and workflow.}
    The \preproc extracts dependency-scoped code metadata and lemma functions from the verification target; the \comp builds lemma and Verus knowledge; the \prv uses requirement analysis and lemma retrieval to generate proof code; and the \san uses Verus diagnostics and toolchain knowledge to repair failed proofs.}
    \label{fig:workflow}
\end{figure*}

\autoref{fig:workflow} shows the automated core of \ourtool.
Given a verification target and Verus documentation, \ourtool constructs three kinds of knowledge: \codeknow{code metadata}, \lemmaknow{lemma knowledge}, and \verusknow{Verus knowledge}.
The \prv then uses this knowledge to generate candidate proof code, and the \san iteratively repairs failed attempts using verifier diagnostics.
The automated workflow ends at verifier-passing code; in the practical workflow, this code is treated as a reviewable patch before upstream integration.

\subsection{\preproc: Code and Lemma Preprocessing}
\label{sec:preproc}

The \preproc analyzes the source code of the target verification library to construct a metadata dependency graph and to extract lemma functions for later comprehension, retrieval, and proof synthesis.
This module corresponds to the upper-left part of \autoref{fig:workflow}.

\textbf{Metadata dependency graph.}
\ourtool builds a typed directed graph over Verus/Rust language objects.
Each node is identified by its fully-qualified path and may represent a function, spec function, lemma function, structure, enum, trait, or type alias.
A directed edge $u \to v$ indicates that verifying or type-checking $u$ may require information about $v$.
Edges capture two families of dependencies.
Type and structure dependencies include signature type references, composite containment, trait/implementation relations, and type-alias relations.
Call and specification dependencies include function calls, spec-function calls, and references appearing in \cc{requires}, \cc{ensures}, and invariants.

\textbf{Dependent code.}
Given an unproven target function $f$, \ourtool defines \codeknow{dependent code} as the language objects collected by traversing outgoing edges from $f$ in the metadata dependency graph.
In our implementation, we use a maximum traversal depth of 3, which covers most cross-file reference chains in the target repositories while keeping the prompt context compact.
The collected objects are deduplicated by fully-qualified path and serialized into prompt-ready snippets containing signatures, specifications, declarations, comments, and module paths.
This gives the LLM verifier-relevant context without indiscriminately including unrelated repository code.

\textbf{Lemma function extraction.}
Lemma functions in Verus do not execute at runtime; they encode logical facts, invariants, and proof strategies used by other proofs.
\ourtool extracts each lemma's name, parameters, preconditions, postconditions, comments, and module location.
These extracted lemma records are passed to the \comp to construct semantic descriptions and to the \prv to support lemma invocation during proof generation.

\begin{figure*}[t]
    \noindent
    \begin{minipage}[t]{0.495\linewidth}
        \input{pygcode/lemmas-left.pygtex}
    \end{minipage}\hfill
    \begin{minipage}[t]{0.495\linewidth}
        \input{pygcode/lemmas-right.pygtex}
    \end{minipage}
    \coderule
    \begin{lstlisting}[caption = {\textbf{Function signature, comment, and specification of lemma functions from Verus and Asterinas.} The highlighted comments are generated by \ourtool.}, label={code:lemmas}]
    \end{lstlisting}
\end{figure*}

\subsection{\comp: Knowledge Comprehension and Synthesis}
\label{sec:comp}

The \comp constructs the knowledge base shown in the middle of \autoref{fig:workflow}.
It turns extracted repository artifacts and Verus documentation into two retrieval layers: \lemmaknow{lemma knowledge} for reusable proof facts and \verusknow{Verus knowledge} for version-aligned verifier guidance.

\textbf{Natural-language descriptions for lemmas.}
RAG over raw source code is brittle because useful lemmas may not share obvious tokens with the current proof goal.
This problem is acute in Verus repositories: standard-library lemmas often include natural-language comments, but project-specific lemmas frequently expose their purpose only through names, signatures, \cc{requires}, and \cc{ensures}.
For each extracted lemma, \ourtool stores its formal signature and specification.
If documentation is missing or sparse, \ourtool asks the LLM to synthesize a concise natural-language description from the lemma's formal interface.
The resulting descriptions are vectorized and stored together with existing comments and documentation, forming the lemma knowledge database used by the \prv.

\autoref{code:lemmas} illustrates this process.
The Verus-library lemma includes a human-written comment, while the CortenMM lemma has no descriptive annotation.
\ourtool synthesizes the highlighted description so that both lemmas can be retrieved through the same semantic interface.

\textbf{Auto-updated Verus knowledge.}
Verus evolves quickly, and proof idioms embedded in static prompts can become stale.
For example, recent Verus versions require explicit \cc{decreases} clauses in cases where older examples did not, which affects benchmark proofs as shown in \autoref{tab:evolution}.
\ourtool therefore parses official Verus documentation and examples into a Verus knowledge database indexed by topic, such as loop invariants, arithmetic overflow, bit-vector reasoning, and specification syntax.
When documentation lags behind toolchain changes, \ourtool can temporarily register relevant Verus pull requests as version-specific knowledge until official documentation catches up.

\subsection{\prv: Requirement-Aware Proof Generation}
\label{sec:prover}

The \prv corresponds to the upper-right part of \autoref{fig:workflow}.
It consumes the target source code, dependent code, and knowledge-base entries to generate candidate proof code.

\textbf{Requirement analyzer.}
Before proof generation, \ourtool asks the LLM to analyze the target function together with its dependent code and summarize the proof requirements.
This requirement analysis is used to query the lemma knowledge database for candidate lemmas that match the current proof obligation.
The retrieved lemma candidates include signatures, preconditions, postconditions, locations, and semantic descriptions.

\textbf{Proof generator.}
The proof generator constructs a structured prompt containing the target function, dependency-scoped code context, retrieved lemmas, and relevant Verus knowledge.
The LLM then generates candidate proof code, which may include assertions, lemma invocations, loop invariants, or auxiliary proof blocks.
The generated proof is not trusted by \ourtool, it is immediately checked by the Verus verifier.

\begin{figure}[t]
    \centering
    \includegraphics[width=0.9\columnwidth]{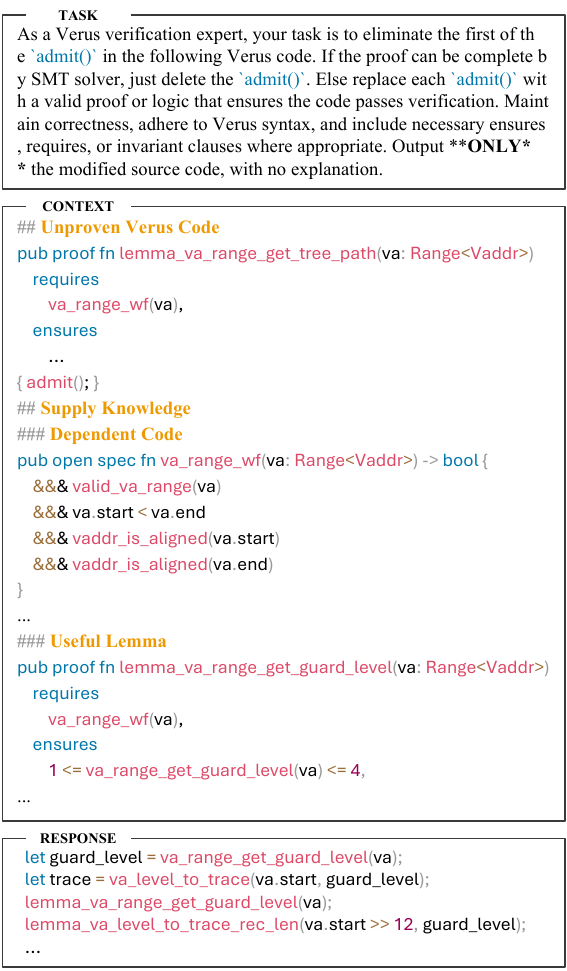}
    \caption{\textbf{Proof-generation prompt for the running example.}
    The prompt combines the target, dependent code, and retrieved lemmas. The response excerpts the generated proof.}
    \label{fig:prompt-prove-main}
\end{figure}

For this target, semantic retrieval supplies lemmas for guard bounds, path length, and node validity. \autoref{fig:prompt-prove-main} shows the final payload and response.

\begin{figure*}[t]
    \centering
    \includegraphics[width=0.9\textwidth]{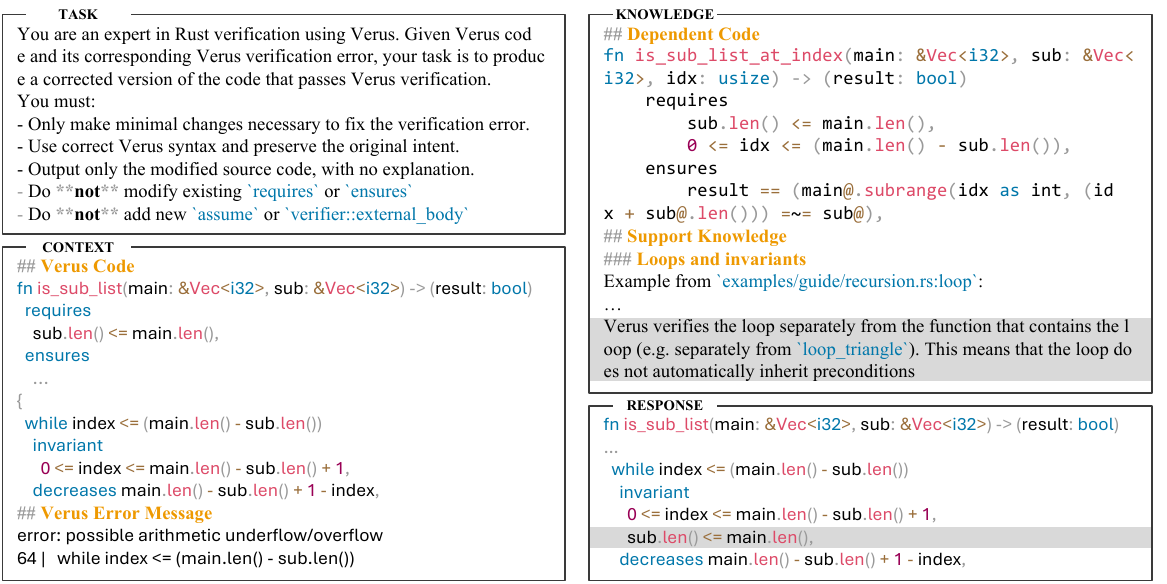}
    \caption{\textbf{Example refinement prompt for \cc{is\_sub\_list}.} 
    We highlight the invariant introduced during the refinement phase and its corresponding supporting knowledge.}
    \label{fig:prompt-refine}
\end{figure*}

\subsection{\san: Error-Driven Proof Refinement}
\label{sec:refiner}

The \san corresponds to the lower-right part of \autoref{fig:workflow}.
If Verus accepts the generated proof, \ourtool outputs verifier-passing code.
If verification fails, \ourtool uses the verifier diagnostics to guide repair.

\textbf{Error triager and fixer.}
It is impractical to include all Verus-related knowledge in every refinement prompt.
Instead, \ourtool classifies the verifier error using the diagnostic text and source location, then retrieves the corresponding Verus knowledge.
The current triager recognizes categories such as loop invariants, arithmetic overflow, bit-vector reasoning, and specification syntax.
The error fixer combines the failed proof, the Verus error message, and the retrieved guidance into a repair prompt for the LLM.

Rather than retrying with a generic instruction, \ourtool supplies error-corresponding Verus knowledge so the LLM can repair the specific verifier failure.

\autoref{fig:prompt-refine} shows a complete refinement example from Verus-Bench.
The initial function requires \cc{sub.len() <= main.len()}, but the generated loop invariant does not preserve this relation.
Consequently, Verus reports a possible arithmetic underflow/overflow at \cc{main.len() - sub.len()} in the loop guard.
Based on both the diagnostic and its loop location, the triager classifies the failure under \emph{Loops and invariants} and retrieves the corresponding Verus guidance: loops are checked separately and do not automatically inherit function preconditions.
The error fixer combines this guidance with the failed code and diagnostic in a repair prompt that requests minimal changes and forbids modifying \cc{requires}/\cc{ensures} or adding assumptions.
The LLM response adds \cc{sub.len() <= main.len()} to the loop invariant, exposing the fact needed to establish that the subtraction is well-defined.

\textbf{New lemmas.}
During refinement, the LLM may identify an auxiliary lemma needed to discharge the current proof.
When this happens, \ourtool adds the new lemma candidate to the generated proof and, if it passes Verus checking and review, incorporates it into the lemma knowledge base for later tasks.

\subsection{Human Review and Workflow Integration}

The automated workflow in \autoref{fig:workflow} ends at verifier-passing code, but repository-scale verification does not end there.
\ourtool treats verifier-passing proof code as a reviewable patch rather than as a trusted final artifact.
The verifier checks formal validity with respect to the written specification, while human reviewers assess whether the specification is appropriate and whether the proof is readable, maintainable, and consistent with project style.
For CortenMM, the practical path is therefore: unverified Verus function, dependency-scoped context extraction, lemma and Verus knowledge retrieval, LLM proof generation, Verus checking and refinement, human review, and upstream-accepted proof code.

\subsection{Implementation}

\ourtool is built on verus-analyzer~\cite{VerusAnalyzer} and LangChain~\cite{Chase2022LangChain}. It comprises about 5,837 lines of Python and 1,015 lines of Rust.
The verus-analyzer front end provides resolved symbols, type information, source ranges, and comments for building the repository-level metadata database.
For Verus features that are newly introduced but not yet reflected in official documentation, \ourtool supports manually registering the corresponding pull requests as temporary Verus knowledge until the documentation catches up.

\section{Experiments}
\label{sec:eval}

\ourtool is evaluated through comparison experiments, ablation studies, and a toolchain-evolution study.
We organize the evaluation around the following research questions:

\begin{itemize}[leftmargin=*]
    \item \textbf{RQ1: Repository-level effectiveness.} How effective is \ourtool on repository-level benchmarks with cross-file dependencies?
    \item \textbf{RQ2: Single-file effectiveness.} How effective is \ourtool on single-file Verus proof-generation benchmarks compared with existing LLM-based tools?
    \item \textbf{RQ3: Knowledge contribution.} How much do code metadata knowledge, lemma knowledge, and Verus knowledge contribute to proof-generation success?
    \item \textbf{RQ4: Toolchain-evolution robustness.} Does version-aligned Verus knowledge improve robustness across Verus releases?
    \item \textbf{RQ5: Practical integration.} Can verifier-passing generated proofs fit into a real review-and-merge workflow? We answer this question through the CortenMM deployment experience in \autoref{sec:practical}.
\end{itemize}

\noindent \textbf{Experiment Setup.}
All experiments were conducted on a server equipped with two Intel Xeon Gold 6230R CPUs (52 cores each, 2.10 GHz), 128 GB of RAM, and Ubuntu 24.04.1 LTS (64-bit).
We configured \ourtool to use the \cc{Claude Sonnet 4.0}~\cite{anthropic2025claude4sonnet} model.
The temperature parameter was fixed at 0.5 for all LLMs throughout the experiments.
The maximum number of refinement queries was set to 10.
Unless otherwise noted, we used Verus \cc{20250813} with default settings.
Each experiment was repeated three times, and we report the union of results across repetitions.

\noindent \textbf{Comparison Tools.}
We compared \ourtool with \autoverus~\cite{yang2024autoverus} and \alphaverus~\cite{aggarwal2024alphaverus}.
To ensure a fair comparison, we configured all tools to use the same underlying LLM, \cc{Claude Sonnet 4.0}.
Because \alphaverus does not release its prompts,\footnote{We requested the prompts from the \alphaverus authors but did not receive a response.} we developed a prompt for it.
We also use \cc{Claude Sonnet 4.0} with 10 refinement queries as a multi-round prompting baseline.
\safe~\cite{chen2024automated} was excluded because it has not released its source code or models.
\autoverus and \alphaverus are included in the single-file comparison, which matches the setting targeted by their released workflows.
For repository-level benchmarks, their workflows do not construct cross-file dependency context or retrieve project lemmas across modules; adding these mechanisms would change the baselines rather than evaluate the original tools.

\noindent \textbf{Benchmarks.}
We evaluate \ourtool on three single-file benchmarks: Verus-Bench, MBPP, and Human-Eval.
We also evaluate it on three repository-level benchmarks: MathSpec-Bench, Memory Allocator, and CortenMM.
Verus-Bench was introduced by \autoverus, whereas MBPP and Human-Eval were used by \alphaverus.
MathSpec-Bench\footnote{\url{https://github.com/rikosellic/verus-mathspec-bench}} is derived by translating mathematical developments from Lean 4 into Verus and is designed to stress lemma reuse and cross-module reasoning.
Memory Allocator and CortenMM provide repository-level verification targets with non-trivial cross-file dependencies.
In the benchmark experiments, we remove the existing proofs for target functions while keeping shared lemmas intact, and require each tool to generate the missing proofs.

\subsection{RQ1: Repository-Level Effectiveness}
\label{sec:eval:rq1}

This experiment evaluates whether \ourtool can handle benchmark tasks with cross-file dependencies.
We apply \ourtool to three repository-level benchmarks: MathSpec-Bench, Memory Allocator, and CortenMM.
For this reason, the repository-level comparison uses the multi-round prompting baseline as the direct prompt-only baseline.
This baseline receives iterative Verus feedback but does not use \ourtool's metadata graph, semantic lemma knowledge, or version-aligned Verus knowledge.

\begin{table}[t]
\centering
\caption{
    \textbf{Proof success on repository-level benchmarks.}
}
\label{tab:repo-level}
\begin{tabular}{lrrr}
\toprule
\textbf{Benchmark} & \textbf{\#Task} & \textbf{\ourtool} & \textbf{Baseline} \\ \midrule
MathSpec-Bench   & 104 & 81 & 10 \\
Memory Allocator & 89  & 50 & 1 \\
CortenMM         & 166 & 52 & 5 
\\
\bottomrule
\end{tabular}
\end{table}

\autoref{tab:repo-level} reports proof success for \ourtool and the multi-round prompting baseline.
Across the three repository-level benchmarks, \ourtool verifies 183 of 359 tasks, achieving a 51.0\% success rate.
The baseline verifies only 16 tasks, or 4.5\%.
On MathSpec-Bench, \ourtool verifies 77.9\% of tasks, compared with 9.6\% for the baseline.
On Memory Allocator, \ourtool verifies 56.2\% of tasks, compared with 1.1\% for the baseline.
On CortenMM, \ourtool verifies 31.3\% of tasks, compared with 3.0\% for the baseline.
\ourtool uses 2,961,245 tokens on MathSpec-Bench, 1,690,465 on Memory Allocator, and 5,041,957 on CortenMM.


\begin{table}[t]
    \centering

\centering
\caption{
    \textbf{Per-task statistics comparison between the single-file (Verus-Bench) and repository-level (CortenMM) benchmarks.}
}
\label{tab:bench-detail}
\begin{tabular}{lrr}
\toprule
\textbf{Statistic}  & Verus-Bench & CortenMM \\ \midrule
{Spec LoC}   & 8           & 193      \\
{Proof LoC}  & 10          & 195      \\
{\#Loop}     & 1.6         & 0.02     \\
{\#Lemma}    & 0.07        & 13.9           
\\ \bottomrule
\end{tabular}
\end{table}

\autoref{tab:bench-detail} explains why repository-level tasks differ from single-file tasks.
Compared with Verus-Bench, CortenMM contains substantially larger specifications and proof bodies, and requires far more lemma invocations per task.
This confirms that repository-scale proof generation is primarily a structural retrieval and lemma-reuse problem, not merely a larger version of single-file proof completion.

\noindent\textbf{Answer to RQ1.}
Repository-level Verus proof generation requires cross-file context and semantic lemma reuse; repeated prompting is insufficient.

\subsection{RQ2: Single-File Effectiveness}
\label{sec:eval:rq2}

\begin{table}[t]
\centering
\caption{
    \textbf{Proof success on single-file benchmarks.}
}
\label{tab:autoverus}

\setlength{\tabcolsep}{3pt}
\resizebox{\columnwidth}{!}{
\begin{tabular}{lrrrrr}
\toprule
\textbf{Benchmark} &
  \textbf{\#Task} &
  \textbf{\ourtool} &
  \textbf{\autoverus} &
  \textbf{\alphaverus} &
  \textbf{Baseline} \\ \midrule
Verus-Bench  & 150 & \textbf{132} & 87  & 53 & 89 \\
MBPP         & 78  & \textbf{65}  & 53  & 21 & 45 \\
Human-Eval   & 85  & \textbf{54}  & 38  & 4  & 5  \\ \midrule
\textbf{SUM} & 313 & \textbf{251} & 178 & 78 & 139 \\
\bottomrule
\end{tabular}
}
\end{table}

\autoref{tab:autoverus} compares proof success across three single-file benchmarks.
Across the three benchmarks, \ourtool verifies 251 of 313 tasks (80.2\%).
This outperforms \autoverus, which verifies 178 tasks (56.9\%), and \alphaverus, which verifies 78 tasks (24.9\%).
Compared with \autoverus, \ourtool verifies 23.3 percentage points more tasks while using 43.2\% of the tokens.
The multi-round prompting baseline verifies 139 tasks (44.4\%), showing that repeated prompting alone is insufficient even in single-file settings.
The evaluation records 5,696,358 tokens and 1,912 LLM queries for \ourtool, compared with 13,498,379 tokens and 3,922 queries.

\noindent\textbf{Answer to RQ2.}
\ourtool outperforms existing LLM-based Verus proof-generation methods on single-file benchmarks while using fewer tokens than \autoverus, showing that its repository-oriented design also benefits local proof tasks.

\subsection{RQ3: Knowledge Contribution}
\label{sec:eval:rq3}

\begin{table}[t]
\centering
\caption{
    \textbf{Proof success for the ablation study.}
}
\label{tab:ablation}

\setlength{\tabcolsep}{3pt}
\resizebox{\columnwidth}{!}{
\begin{tabular}{lrrrrr}
\toprule
\textbf{Benchmark} & \textbf{\#Task} & \textbf{\ourtool} & \textbf{w/o Code} & \textbf{w/o Lemma} & \textbf{w/o Verus} \\ \midrule
Verus-Bench      & 150 & 110 & 110 & 108 & 88 \\
MBPP             & 78  & 65  & 66  & 64  & 46 \\
Human-Eval       & 85  & 54  & 53  & 50  & 4  \\
MathSpec-Bench   & 104 & 81  & 42  & 63  & 75 \\
Memory Allocator & 89  & 50  & 50  & 51  & 2  \\
CortenMM         & 166 & 52  & 10  & 7   & 22 \\

\bottomrule
\end{tabular}
}

\end{table}

\autoref{tab:ablation} summarizes the ablation results for code metadata knowledge, lemma knowledge, and Verus knowledge.
On the three single-file benchmarks, removing Verus knowledge causes the largest degradation, especially on Human-Eval (54 to 4 proved tasks), showing that version-aligned Verus guidance is critical for syntax and proof-obligation handling.
Removing code metadata or lemma knowledge has limited impact on single-file tasks, where most relevant context is already local.

The trend reverses on repository-level benchmarks.
Removing code metadata knowledge reduces success from 81 to 42 on MathSpec-Bench and from 52 to 10 on CortenMM.
Removing lemma knowledge also causes large drops, from 81 to 63 on MathSpec-Bench and from 52 to 7 on CortenMM.
These results show that structural project knowledge and lemma reuse are the main drivers of repository-level scalability.

\noindent\textbf{Answer to RQ3.}
Verus knowledge is most important for single-file robustness, while code metadata and lemma knowledge are essential for repository-scale verification.

\subsection{RQ4: Robustness to Toolchain Evolution}
\label{sec:eval:rq4}

\begin{table}[t]
\centering
\caption{
    \textbf{Evolution experiment on single-file benchmarks.}
    \cc{20250813} is the latest Verus release used in our evaluation.
    K for \ourtool and A for \autoverus.
}
\label{tab:evolution}

\begin{tabular}{lcccccc}
\toprule
\multirow{2}{*}{\textbf{Benchmark}} & \multicolumn{2}{c}{\cc{20250813} (latest)} & \multicolumn{2}{c}{\cc{20250630}} & \multicolumn{2}{c}{\cc{20250328}} \\ \cmidrule(lr){2-3} \cmidrule(lr){4-5} \cmidrule(lr){6-7}
             & K & A & K & A & K                & A \\ \midrule
Verus-Bench  & \textbf{132}      & 87         & 108      & 90         & 130      & 129        \\
MBPP         & 65       & 53         & 67       & 57         & \textbf{70}       & 62         \\
Human-Eval   & \textbf{54}       & 38         & 42       & 40         & \textbf{54}       & 43         \\ \midrule
\textbf{SUM} & 251      & 178        & 217      & 187        & \textbf{254}      & 234       \\
\bottomrule
\end{tabular}
\end{table}

\autoref{tab:evolution} compares \ourtool with \autoverus across three Verus releases, from the oldest evaluated release \cc{20250328} to the latest evaluated release \cc{20250813}.
On \cc{20250328}, \ourtool verifies 254 tasks, compared with 234 for \autoverus.
On \cc{20250630}, both tools are affected by a toolchain change, but \ourtool still verifies 217 tasks compared with 187 for \autoverus.
On \cc{20250813}, which introduced a breaking requirement for explicit \cc{decreases} clauses, \autoverus drops to 178 tasks, whereas \ourtool verifies 251 tasks.
Thus, from \cc{20250328} to \cc{20250813}, \autoverus drops by 23.9\%, while \ourtool drops by only 1.2\%.

This result aligns with the CortenMM maintenance experience: long-lived Verus repositories must adapt to verifier and library evolution.
\ourtool addresses this by retrieving version-aligned Verus knowledge during generation and refinement, instead of relying only on static prompt examples.

\noindent\textbf{Answer to RQ4.}
Version-aligned Verus knowledge improves robustness under toolchain evolution, which is necessary for maintaining long-lived Verus repositories.

\section{Practical Experience}
\label{sec:practical}

We applied \ourtool to CortenMM, Asterinas's verified memory-management module.
This setting represents a realistic repository-scale verification workload.
Its proofs draw on page-table specifications, arithmetic lemmas, state-machine invariants, bit-level code, and repository-wide helper definitions.
Given an unverified Verus function, \ourtool extracts dependency-scoped context, retrieves project-specific and Verus-library lemmas, generates proof code, checks it with Verus, refines failures using verifier diagnostics, and leaves the final proof for human review before upstream integration.

\begin{table*}[t]
\centering
\caption{
    \textbf{CortenMM proof patches generated by \ourtool.}
    We group the 23 previously unverified targets by the properties described in \autoref{sec:practical}; all patches passed Verus and were accepted upstream by Asterinas developers.
}
\label{tab:vostd}

\begin{tabular}{lrrr}
\toprule
\textbf{Property} & \textbf{Verified Targets} & \textbf{Proof LoC} & \textbf{Representative Examples} \\
\midrule
Mathematical properties of the tree model & 9 & 332 & \cc{lemma\_get\_child}; tree-path and subtree lemmas \\
State-machine invariants and refinement & 12 & 1,019 & \cc{next\_refines\_next}; lock/unlock transition proofs \\
Functional correctness of executable code & 2 & 36 & \cc{lock\_range}; \cc{page\_size} \\
\midrule
\textbf{Total} & \textbf{23} & \textbf{1,387} &  \\
\bottomrule
\end{tabular}
\end{table*}

\autoref{tab:vostd} summarizes the CortenMM proof patches generated by \ourtool by property class.
Overall, \ourtool generated 1,387 lines of proof code, including 6 newly introduced lemma functions, and verified 23 functions that were previously unverified.
The generated code accounts for 21.0\% of the proof code in CortenMM.
The generated proofs were reviewed and accepted by the Asterinas developers.

\subsection{Verified Properties in CortenMM}
The completed tasks cover three representative classes of properties in the CortenMM verification effort.

\textbf{Mathematical properties of the tree model.}
The page table is modeled as a 4-level, full 512-ary tree, where each node represents a memory page.
Each node is assigned a unique node ID via preorder traversal, and can also be represented by a trace, i.e., the sequence of offsets along the path from the root to the node.
Structural notions such as ancestors, descendants, and subtrees are thus reduced to comparisons between IDs.
Establishing this correspondence, however, requires non-trivial nonlinear arithmetic, which is beyond the direct capabilities of SMT solvers and remains challenging even for human reasoning.
\ourtool handles this complexity by reusing existing lemmas about nonlinear arithmetic.
For example, the \cc{get\_child} function takes a node ID, derives its trace, appends a given offset, and maps it back to an ID.
In \autoref{code:lemma-get-child}, the associated \cc{lemma\_get\_child\_sound} asserts two key facts: the resulting ID lies within the valid range of node IDs, and its parent corresponds to the original node.
\ourtool proves the former by inserting arithmetic assertions and invoking existing lemmas, and the latter by reasoning about traces and sequences.

\begin{figure}[t]
    \input{pygcode/lemma-get-child.pygtex}
    \coderule
    \begin{lstlisting}[caption = {\textbf{\cc{lemma\_get\_child\_sound} from CortenMM.} We highlight the proof code generated by \ourtool.}, label={code:lemma-get-child}]
    \end{lstlisting}
\end{figure}

\textbf{Invariants of high-level system design.}
Moving up from arithmetic reasoning, Verus supports the verification of concurrent systems by encoding system designs as state machines.
\autoref{code:lemma-rcu} illustrates the state machine of cursor operations in memory management, where a cursor locks a range of memory pages.
The user specifies both the state fields and the transitions that manipulate them.
To ensure correctness, the user must also provide invariants that capture key system properties and prove that these invariants are preserved under all transitions.
Such system-level reasoning is an advanced Verus capability and remains beyond the scope of prior automation approaches.
Despite the scarcity of examples of this Verus-specific technique, \ourtool successfully learns its specialized syntax by parsing official documentation and automates the proofs for 10 out of 15 transitions.
Manual inspection reveals that the unprovable transitions expose specification issues rather than missing proof steps.

\begin{figure}[t]
    \noindent
        \input{pygcode/lemma-rcu.pygtex}
    \coderule
    \begin{lstlisting}[caption = {\textbf{\cc{protocol\_unlock\_start\_inductive} and its dependent state fields and transition from CortenMM.} We highlight the proof code generated by \ourtool.}, label={code:lemma-rcu}]
    \end{lstlisting}
\end{figure}

\textbf{Functional correctness of executable code.}
At the implementation level, CortenMM employs bit-level manipulations in its executable code to boost performance, which are notoriously hard to verify.
First, specifications are typically written with exponentials rather than bit-level operators, which eases reasoning but complicates alignment with the implementation.
Second, these tricks depend on the precise coordination of multiple interdependent system parameters.
Third, Verus delegates certain reasoning tasks to external solvers, such as \cc{bit\_vector} and \cc{nonlinear\_arith}, isolated from the surrounding proof context, requiring additional auxiliary assertions at the proof boundary.
\autoref{code:lemma-page-size} shows the \cc{page\_size} function, which returns the size of a memory page at a given level.
Here \cc{C: PagingConstsTrait} provides constants for memory management.
Despite its straightforward design, verifying this function is nontrivial due to the interplay of overflow restrictions, trait and function dependencies, external solver queries, and bit-operation lemmas, all of which \ourtool discharges automatically.

\begin{figure}[t]
    \noindent
        \input{pygcode/lemma-page-size.pygtex}
    \coderule
    \begin{lstlisting}[
        caption = {\textbf{\cc{page\_size} and its specification from CortenMM.} We highlight the proof generated by \ourtool.},
        label={code:lemma-page-size}]
    \end{lstlisting}
\end{figure}

\subsection{Boundary Cases}

\noindent\textbf{Incorrect specifications.}
In CortenMM, not every failed proof attempt reflected a missing proof step.
While verifying the Asterinas RCU state machine, \ourtool repeatedly failed to prove that \cc{protocol\_lock\_skip} preserves the state-machine invariants.
The transition skips a node \cc{nid} and its subtree during locking when the parent node's page-table entry is \cc{void}; conceptually, the operation locks the skipped subtree in one step and advances the cursor to \cc{NodeHelper::next\_outside\_subtree(nid)}.
The original specification enforced the cursor update, the empty-PTE condition, consistency between the removed cursor state and \cc{nid}, and the parent/offset relationship used to access the PTE.
However, the specification omitted two conditions needed for soundness.
First, the cursor must stay within the subtree rooted at \cc{rt}; applying the transition once the cursor reaches \cc{NodeHelper::next\_outside\_subtree(rt)} moves it beyond the legal range and violates \cc{wf\_cursor}.
Second, the transition must require \cc{nid != rt}, because otherwise it attempts to access the root node's parent.
\autoref{code:practical-fail-spec} shows the corrected specification lines added by a human expert.
This case illustrates that verifier failures can expose specification bugs: when the specification is wrong, a proof generator should not fabricate proof code to make the verifier pass.

\begin{figure}[t]
    \input{pygcode/fail-spec.pygtex}
    \coderule
    \begin{lstlisting}[caption = {\textbf{Specification repair for \cc{protocol\_lock\_skip}.} Highlighted lines are the missing conditions added by a human expert after proof attempts exposed the specification error.}, label={code:practical-fail-spec}]
    \end{lstlisting}
\end{figure}

\noindent\textbf{Human proof-code cleanup.}
Even when generated proofs are valid, human experts may still revise them before integration.
In one \cc{initialize\_inductive} proof, \ourtool generated more than 100 lines across seven proof blocks, corresponding to the seven invariants it needed to establish.
A human expert later proved the same obligations with 34 lines across two proof blocks by using a more compact proof structure and relying on expert knowledge about which intermediate assertions were redundant.
This difference does not invalidate the generated proof, but it shows why practical deployment should treat LLM-generated proofs as reviewable patches: the verifier checks correctness with respect to the specification, while human review improves maintainability and style.

\noindent\textbf{Takeaway.}
Verifier-passing generated proofs are useful as reviewable patches, but practical integration still depends on human judgment for specification repair and proof cleanup.

\section{Lessons Learned and Limitations}
\label{sec:discuss}

\noindent\textbf{Lesson 1: Dependency-scoped knowledge is the key scaling unit.}
In CortenMM, proof obligations often depend on definitions, invariants, and helper lemmas spread across page-table specifications, state-machine models, and executable Rust code.
Target-file prompting misses these dependencies, while broad repository prompting introduces noise and cost.
The useful unit for repository-scale proof generation is therefore dependency-scoped knowledge: verifier-relevant code context plus semantically indexed lemmas.
Because many project lemmas expose their role through signatures, preconditions, postconditions, and module paths rather than comments, verification teams should treat dependency metadata and proof-intent documentation as part of the proof infrastructure.

\noindent\textbf{Lesson 2: Proof generation is also proof maintenance.}
The evolution experiments show that Verus/toolchain updates can break previously valid proofs even when the verified program has not changed.
For repository-scale verification, this is an engineering concern rather than a benchmark artifact: proof maintenance must track the verifier version, library APIs, error messages, and accepted proof idioms.
\ourtool mitigates this problem through version-aligned knowledge and verifier-aware refinement, but it does not eliminate the need for continuous validation when projects upgrade Verus.
Long-lived verification projects should pin verifier versions for experiments, archive version-specific proof logs, and run proof CI after toolchain updates.

\noindent\textbf{Lesson 3: Generated proofs should remain in an expert workflow.}
\ourtool is designed to reduce repetitive proof construction, not to replace proof engineers.
In CortenMM, generated proofs were checked by Verus and then reviewed before upstream integration.
This review path matters because Verus checks proof validity against the written specification, while humans judge whether the specification is correct and the proof is maintainable.
Repeated proof failures exposed specification issues in several state-machine transitions, and valid generated proofs still sometimes needed cleanup.
For example, \ourtool produced more than 100 lines for \cc{initialize\_inductive}, while a human expert later compressed the same obligations to 34 lines.
Thus, LLM-generated proof code should enter the same review path as human-written proof patches, with proof minimization and style cleanup before upstream submission.

\noindent\textbf{Limitations.}
First, \ourtool does not validate whether a specification captures the intended system behavior; it can surface suspicious proof failures, but specification repair remains a human responsibility.
Second, verifier-passing proofs may still be unsuitable for integration because the verifier does not assess readability or maintainability.
Finally, our evidence is strongest for Verus-based Rust repositories.
The same knowledge-centric workflow may transfer to other SMT-based verifiers such as Dafny, but theorem provers with explicit proof states would require different representations of context, lemma retrieval, and verifier feedback.

\section{Related Work}
\label{sec:relwk}

\textbf{LLM-Assisted Verification for Rust.}
LLM-assisted proof generation for Rust is still emerging, with Verus as the main target language.
\autoverus~\cite{yang2024autoverus} introduced Verus-Bench and showed that prompt-based generation can automate many single-file Verus proofs.
Its prompts encode expert proof strategies and common proof patterns, but the method has limited access to repository-specific dependencies and reusable project lemmas.
\safe~\cite{chen2024automated} and \alphaverus~\cite{aggarwal2024alphaverus} improve proof generation through stronger refinement mechanisms.
\safe uses GPT-4o-generated data to fine-tune a local model, while \alphaverus uses verifier-guided tree search.
These approaches are effective on constrained tasks, but they do not directly target the workflow bottlenecks we observed in Asterinas/CortenMM: cross-module dependency discovery, project-specific lemma reuse, and Verus/toolchain evolution.
\ourtool is complementary to refinement-heavy methods.
It focuses on constructing the right verification context before generation, then uses verifier feedback for bounded repair.

Domain-specific Rust verification tools take a different route.
OwlC~\cite{singh2025owlc}, for example, generates Rust code and Verus proofs from high-level security protocol specifications.
Such tools can provide strong automation when the domain language captures the target problem, whereas \ourtool targets general Verus proof generation over existing Rust repositories.
Rust and Verus have also been used in manually engineered verification projects for systems software, including verified kernels, security modules, and distributed-system components~\cite{atmosphere,zhou2024verismo,sun2024anvil,li2022linear,hance2023sharding,hance2023leaf}.
These projects demonstrate the expressiveness of modern verification tools, but they also illustrate the amount of repository-specific proof knowledge that automation must recover and reuse.

\noindent\textbf{Proof Generation Beyond Rust.}
LLM-based proof generation has also been studied for theorem provers such as Rocq~\cite{coq}, Isabelle~\cite{nipkow2002isabelle}, and Lean~\cite{moura2021lean}.
These systems expose explicit proof states and tactic languages, allowing tools to predict proof steps and retrieve similar proof states.
Rango~\cite{thompson2024rango}, for instance, retrieves relevant lemmas, proof scripts, and proof states to guide tactic prediction in Rocq.
Verus differs from these systems because users write proof annotations that are translated into SMT queries rather than manipulating an explicit proof state.
As a result, repository structure, available lemmas, verifier diagnostics, and solver behavior become central to automation.

\noindent\textbf{Verified Systems and Program Logics.}
The broader verification literature includes machine-checked kernels, verified compilers, and program logics for concurrent software~\cite{klein2009sel4,gu2016certikos,leroy2016compcert,jung2015iris}.
Our work does not replace these verification foundations; instead, it targets a practical bottleneck that appears when proof engineers must extend and maintain such assurance arguments inside an evolving Rust repository.

Dafny~\cite{leino2010dafny} is closer to Verus because it is also an SMT-based program verifier.
Prior work on Dafny proof automation has used fine-tuning~\cite{poesia2024dafny}, prompt engineering~\cite{misu2024towards}, and proof-localization techniques.
\laurel~\cite{mugnier2025laurel} synthesizes Dafny assertions by identifying insertion locations and retrieving syntactically similar proofs.
These techniques improve standalone proof completion, but they leave open the repository-scale concerns emphasized in this paper: dependency-scoped context construction, semantic lemma retrieval across modules, and robustness to verifier evolution.

\section{Conclusion}
\label{sec:conclusion}

This paper presented \ourtool, a knowledge-centric workflow for LLM-assisted Verus proof generation in repository-scale Rust systems.
Our experience with Asterinas/CortenMM shows that practical proof generation depends on more than local prompt quality: tools must recover cross-module dependencies, reuse project-specific lemmas, adapt to Verus/toolchain evolution, and leave generated proofs in a form that maintainers can review.
\ourtool addresses these needs through dependency-aware context extraction, semantic lemma retrieval, and verifier-aware refinement.

In CortenMM, \ourtool generated 1,387 lines of proof code, including 6 new lemma functions, and verified 23 previously unverified functions, contributing 21.0\% of the proof code.
Across repository-level benchmarks, \ourtool achieved a 51.0\% success rate, and the single-file evaluation shows that the same workflow remains competitive with existing LLM-based proof-generation methods.
These results suggest that LLM-assisted proof generation can move beyond benchmark-only settings and become a practical aid for maintaining verified Rust repositories, provided that generated proofs remain subject to verifier checking and human review.

\begin{acks}
We thank the anonymous reviewers for their insightful comments and feedback.
We are grateful to the Asterinas community for their support in analyzing the correctness of specifications.
This research was supported by Ant Group Postdoctoral Programme.
\end{acks}


\appendix

\section*{Data Availability Statement}
The KVerus GitHub repository is available at \url{https://github.com/asterinas/KVerus}. 
It is also archived at \url{https://doi.org/10.5281/zenodo.19937769}. 
The artifact includes the KVerus source code, reproduction scripts, and prompts. 
The benchmarks used in \autoref{sec:eval} can be downloaded by the provided scripts with fixed repository commits and Verus versions. 
The merged upstream CortenMM PRs are also listed in the artifact.


\bibliographystyle{ACM-Reference-Format}
\bibliography{paper}

\end{document}